\documentstyle[epsf,prl,aps]{revtex}
\begin{document}
\title{Exact Correlation Amplitude for the S=1/2 Heisenberg Antiferromagnetic Chain} 
\author{Ian Affleck}
\address{Department of Physics and Astronomy and Canadian Institute for Advanced
Research, \\ University of British Columbia, Vancouver, BC,
Canada, V6T 1Z1}
 \maketitle \begin{abstract}
The exact amplitude for the asymptotic correlation function in the S=1/2 Heisenberg 
antiferromagnetic chain is determined:
$$<S^a_0S^b_r>\to  (-1)^r\delta^{ab}(\ln r)^{1/2}/[(2\pi )^{3/2}r].$$  The behaviour 
of the correlation functions for small xxz anisotropy  and the form of finite-size 
corrections to the correlation function are also analysed.
\end{abstract}

The asympototic behaviour of the equal-time correlation function in the Heisenberg 
antiferromagnetic chain has been difficult to determine 
numerically\cite{Kubo,Liang,Sandvik,Hallberg,Koma} because of the presence of a 
marginally irrelevant operator.  This leads\cite{Affleck,Singh} to a logarithmic factor 
of $\sqrt{\ln r}$ and also to finite size effects which only vanish as $1/\ln L$ where L is 
the system size.  This marginal operator leads to logarithmic corrections, sometimes 
multiplicative and sometimes additive, to most long distance, low energy  properties of 
the model.  In particular  recent experiments on Sr$_2$CuO$_3$ found evidence for the 
predicted \cite{Eggert}  logarithmic additive correction to the susceptibility \cite{Sr}.

On the other hand, logarithmic corrections are absent for the xxz model, with 
Hamiltonian:
\begin{equation} H = \sum_i[(S^x_iS^x_{i+1}+S^y_iS^y_{i+1}+
\gamma S^z_iS^z_{i+1}],\end{equation}
for $\gamma \neq 1$.
The model exhibits critical behaviour for $-1\leq \gamma \leq 1$, with asympotic 
correlation functions:
\begin{eqnarray}
G^x(r)\equiv <S^x_0S^x_r>&\to& (-1)^rA_xr^{-\eta}\nonumber \\
G^z(r)\equiv <S^z_0S^z_r>&\to& (-1)^rA_zr^{-1/\eta},\label{exp}\end{eqnarray}
with
\begin{equation} \eta = 1-[\cos^{-1}\gamma]/\pi \ \  (0\leq \eta \leq 1)
.\label{eta}\end{equation}
What appears to be an exact formula for the amplitude $A_x(\gamma )$ was recently 
conjectured \cite{Lukyanov}:
\begin{equation}A_x(\gamma ) = {(1+\xi )^2\over 8}\left[{\Gamma ({\xi \over 
2})\over 2\sqrt{\pi}\Gamma \left({1\over 2}+{\xi \over 2}\right)}\right]^\eta \times 
\exp \left\{-\int_0^\infty {dt\over t}\left({\sinh (\eta t)\over \sinh (t) \cosh[(1-\eta )t]}-
\eta e^{-2t}\right)\right\}.\label{exact}\end{equation}
Here:
\begin{equation} \xi \equiv {\eta \over 1-\eta}.\end{equation}

The main purpose of the present report is to determine the exact amplitude in the 
logarithmic, xxx case, $\gamma =1$, giving the result in the abstract.  To do so it will 
be neccessary to consider the form of these correlation functions for $\gamma$ only 
slightly less than 1 where a crossover from logarithmic to non-logarithimic behaviour 
occurs.  The other amplitude, $A_z$, is not known in general.  We will show that:
\begin{equation} \lim_{\gamma \to 1^-}A_z/A_x=4.\end{equation}
The order of limits here is crucial; right at $\gamma =1$ the amplitudes of the 
(logarithmic) correlation functions $G^x$ and $G^z$ are equal. We also discuss the 
form of finite-size corrections for the correlation function on a ring of length L with 
periodic boundary conditions, $G(r,L)$. 

\begin{figure}
\epsfxsize=10 cm
\centerline{\epsffile{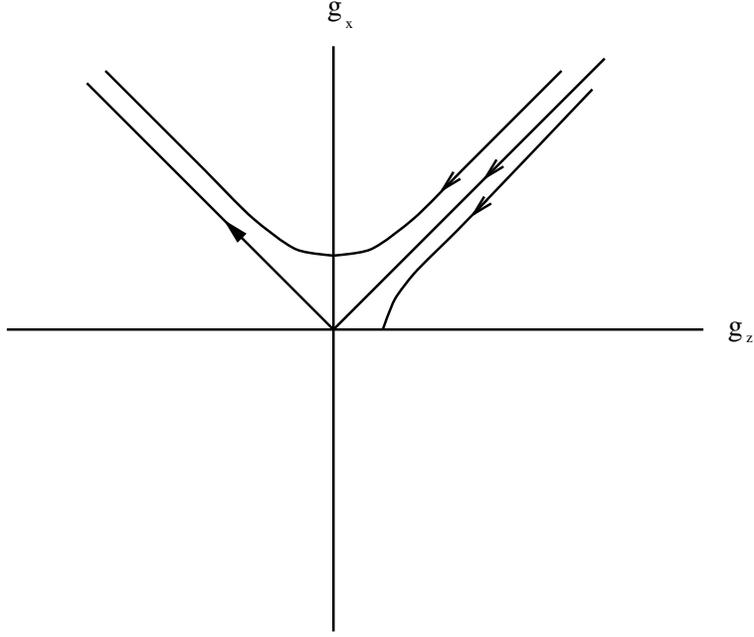}}
\caption{The Kosterlitz-Thouless RG flows of Eq. (\protect{\ref{beta}}).}
\label{fig:KT}
\end{figure}

The subsequent calculations are based on the continuum limit bosonized approximation 
to the xxz model.  We follow the notation of \cite{Affleck}.  The Hamiltonian density 
may be written:
\begin{equation}
{\cal H} = {\cal H}_0-
(8\pi^2/\sqrt{3})[g^x(J_L^xJ_R^x+J_L^yJ_R^y)+g^z(J_L^zJ_R^z)].\label{Hbos}
\end{equation}
Here ${\cal H}_0$ is the Hamiltonian density for a free boson, of compactification 
radius $R=1/\sqrt{2\pi}$, or equivalently, the SU(2) level 1 Wess-Zumino-Witten 
(WZW) non-linear $\sigma$ model.  $\vec J_{L,R}$ are the left and right moving 
currents.  (We set the spin-wave velocity equal to 1.) For the isotropic model, with 
$\gamma =1$, $g^x=g^z=g$ is of O(1).  The rather cumbersome normalization in Eq. 
(\ref{Hbos}) is dictated by the convention that the operator multiplying $g$ in the 
isotropic case have a correlation function with unit amplitude.
For the xxz model with $\gamma$ close to 1,
\begin{equation} g^z-g^x\propto 1-\gamma .\end{equation}  These coupling constants 
obey the Kosterlitz-Thouless renormalization group (RG) equations:
\begin{eqnarray}
\beta_z\equiv dg_z/d \ln L&=&-(4\pi /\sqrt{3})g_x^2,\nonumber \\
\beta_x\equiv dg_x/d\ln L &=& -(4\pi /\sqrt{3})g_xg_z.\label{beta}\end{eqnarray}
The RG trajectories are sketched in Fig. \ref{fig:KT}.  
$g_z^2(L)-g_x^2(L)$ is an RG invariant along the flow.  For $g_z>|g_x|$, the flow is to 
a fixed line, the positive $g_z$ axis.
  $g_z^2(L)-g_x^2(L)=g_z^2(\infty )$ along these trajectories.  
Using the abelian bosonization formula 
$J^z_L=-(1/\sqrt{8\pi})(\partial_0+\partial_1)\phi$, we find that, at the
fixed point, the effective Lagrangian is:
\begin{equation}{\cal L}=(1/2)(\partial_{\mu}\phi )^2 [1-(2\pi g_z(\infty
)/\sqrt{3}].\label{Leff}\end{equation}  
The staggered part of the local spin operators may be written in non-abelian 
bosonization notation as:
\begin{equation} \vec S_i\propto (-1)^i \hbox{tr}g \vec \sigma ,\end{equation}
where $g$ is the two dimensional unitary  matrix field of the WZW model.
In terms of abelian bosonization:
\begin{eqnarray}
S^z_i&\propto& (-1)^i \sin (\phi /R)\nonumber \\
S^x_i&\propto& (-1)^i \cos (2\pi R\tilde \phi ),
\label{bos}\end{eqnarray}
where $\tilde \phi$ denotes the dual field and $R=1/\sqrt{2\pi}$.  From Eqs. 
(\ref{Leff}) and (\ref{bos}) we can determine the correlation exponents of Eq. 
(\ref{exp}) with:
\begin{equation} \eta = 1-2\pi g_z(\infty )/\sqrt{3}.\end{equation}
Note that, using Eq. (\ref{eta}), determined from the Bethe ansatz solution, the value of 
$g_z(\infty )$ is determined exactly.  The scaling dimensions of the staggered spin 
operators tr$g\sigma^x$ and tr$g\sigma^z$ are given by $\eta /2$ and $1/2\eta$ 
respectively. To study the logarithmic behaviour, we will also need the anomalous 
dimensions for small non-zero $g_i$, along the RG trajectories.  These
 can be determined from the 3-point Green's functions
$<\hbox{tr}g\sigma^aJ_L^bJ_R^b\hbox{tr}g\sigma^a>$ as in \cite{Affleck}.  Using 
the fact that the operator product expansion gives:
\begin{equation} J_L^b(z)J_R^b(\bar z)g(z',\bar z')\to {(1/4)\sigma^bg\sigma^b\over 
|2\pi (z-z')|^2}+\ldots ,\end{equation}
we conclude that:
\begin{equation} 
<\hbox{tr}g\sigma^aJ_L^bJ_R^b\hbox{tr}g\sigma^a>\propto \hbox{tr} 
(\sigma^a\sigma^b\sigma^a\sigma^b)=4\delta^{ab}-2.\end{equation}
Thus, to linear order, the conclusion is:
\begin{eqnarray} \gamma_x&=&1/2-(\pi/\sqrt{3})g_z,\nonumber \\
\gamma_z&=&1/2+(\pi/\sqrt{3})(g_z-2g_x).\end{eqnarray}

In discussing the asympototic correlation functions, using bosonization, it is convenient 
to introduce uniform and staggered terms:
\begin{equation} G^i(r)\to G^i_u(r)+(-1)^rG^i_s(r),\end{equation}
where $G^i_u$ and $G^i_s$ vary slowly on the scale of a lattice spacing.  These two 
terms correspond to different Green's functions in the continuum limit field theory.  In 
this paper we only discuss the staggered term.

The staggered correlation functions (for an infinite spin chain) obey the RG equations:
\begin{equation}
[\partial /\partial \ln r + \sum_j\beta_j (\vec g)\partial /\partial g_j+2\gamma_i (\vec g)] 
G_s^i(r,\vec g)=0.\label{RG}\end{equation}  Here $\vec g\equiv (g_z,g_x)$. This 
follows from the fact that a rescaling of the length is equivalent to a change in the 
values of the effective coupling constants together with a rescaling of the fields with 
exponents $\gamma_i$. The solution of Eq. (\ref{RG}) is:
\begin{equation} G_s^i(r,\vec g^0)=\exp \{-2\int_{r_0}^r d\ln r' \gamma_i[\vec g( 
r')]\}F^i[\vec g(r)],\label{corr}\end{equation}
where the $F^i$ are arbitrary functions of $\vec g(r)$, the solution of the RG equations,
\begin{equation} dg_i/d\ln r = \beta_i(\vec g).\end{equation}
Here $\vec g^0\equiv \vec g(r_0)$, denotes the value of the ``bare'' couplings at some 
reference ``ultraviolet cut off'' scale $r_0$ of order a lattice spacing.  Since, for large r, 
$g_x(r)<<1$, we may expand the functions $F_i[\vec g(r)]$ perturbatively in $g_x(r)$. 
Exactly this procedure is used to analyse deep inelastic scattering data in quantum 
chromodynamics.  It is known as ``renormalization group improved perturbation 
theory''.  To lowest order these functions are just constants. 
Integrating the RG equations for the effective coupling constants, Eq. (\ref{beta}) we 
obtain:
\begin{eqnarray} g_x(r)&=&{\sqrt{3}\epsilon \over 4\pi}\hbox{cosech} (\epsilon \ln 
r)\nonumber \\
g_z(r)&=&{\sqrt{3}\epsilon \over 4\pi}\coth (\epsilon \ln r),\end{eqnarray}
where we have defined:
\begin{equation} \epsilon \equiv 2(1-\eta) = 4\pi g_z(\infty )/\sqrt{3}.\end{equation}
Now performing the integration over $\gamma_i(\ln r)$ in Eq. (\ref{corr}), we obtain:
\begin{eqnarray}
 G_s^x(r)&\to&  {A_x\over r^{1-\epsilon /2}}(1-r^{-2\epsilon })^{1/2}
\nonumber \\
 G_s^z(r) &\to&  {A_z\over r^{1+\epsilon /2}}{(1-r^{-\epsilon})^{1/2}\over
(1+r^{-\epsilon})^{3/2}}.\end{eqnarray}
Note that we have defined the normalization constants so that the asympotic large-r 
behaviour is as in Eq. (\ref{exp}).  Also note that, for $\epsilon <<1$, both correlation 
functions exhibit logarithimic behaviour over an intermediate range of r, $1<<\ln 
r<<1/\epsilon $.  In this range of r, we obtain:
\begin{eqnarray}  G_s^x&\approx&  \sqrt{2\epsilon} A_x{(\ln r)^{1/2}\over 
r}\nonumber \\
 G_s^z(r)&\approx&  {\sqrt{\epsilon}A_z\over 2^{3/2}}{(\ln r)^{1/2}\over 
r}.\end{eqnarray}

Now consider taking the limit $\epsilon \to 0$, corresponding to the isotropic 
Heisenberg antiferromagnet.  We see that in order for the correlation functions to 
remain finite at fixed r as $\epsilon \to 0$ we must have $A_x\propto 1/\sqrt{\epsilon}$.  
Furthermore, in order to obtain the isotropic result, $ G_s^x(r)= G_s^z(r)$, we must 
have $A_z/A_x\to 4$, as $\epsilon \to 0$.  Thus for small but finite $\epsilon$, $ 
G_s^x(r)\approx  G_s^z(r)$ in the intermediate range of r, but at very large r they 
exhibit slightly different exponents and amplitudes differing by a factor of 4.

The exact amplitude, $A_x(\eta )$ of Eq. (\ref{exact}) can be evaluated in closed form 
in the limit $\eta \to 1$, $\epsilon \to 0$.  In this limit we may approximate $\sinh \eta 
t/\sinh t\approx e^{-\epsilon t/2}$ in the first term of the integrand and $\eta \approx 1$ 
in the second term.  The integral can then be done exactly, giving:
\begin{equation} A_x\to {1\over 4(\epsilon)^{1/2}\pi^{3/2}}.\end{equation}
This diverges as $1/\sqrt{\epsilon}$, as expected.
Thus we conclude, in the isotropic case:
\begin{equation}  G_s^z(r)= G_s^x(r)\to  {1\over (2\pi )^{3/2}}{(\ln r)^{1/2}\over 
r}.\end{equation}
The asymptotic form of the Fourier transform for $k\approx \pi$ is thus given by:
\begin{equation} G(k)\equiv \sum_{r=-\infty}^\infty G(r)e^{ikr}\to  {4\over 3(2\pi 
)^{3/2}}|\ln |k-\pi ||^{3/2}.\label{Gk}\end{equation}
Note that the effect of the $(\ln r)^{1/2}$ factor is to change the power of $|\ln |k-\pi ||$ 
from 1 to 3/2.  If such a weak singularity could be observed, this formula might be 
useful to check the normalization in neutron scattering experiments.  It follows from the 
above analysis that, for small xxz anisotropy, this isotropic formula remains valid down 
to exponentially small values of $k-\pi$, making the log singularity of Eq. (\ref{Gk}) 
observable.

Several efforts have been made to check the field theory prediction of logarithmic 
behaviour numerically \cite{Kubo,Liang,Sandvik,Hallberg,Koma}.  Hallberg et al. 
\cite{Hallberg}  obtained the above asympototic behaviour but with an amplitude of 
.06789 in place of the exact result $(2\pi)^{-3/2} = .06349364 \ldots$. This result was 
obtained from density matrix renormalization group calculations on rings of up to 70 
sites using finite-size extrapolation.  Koma and Mitzukoshi \cite{Koma} also obtained 
the above form with an amplitude of .065. [Alternatively, if they let the power of the 
logarithm be a free parameter they obtained a slightly better fit with a power of .47 
instead of 1/2 and an amplitude of .071.]  This was obtained using exact diagonalization 
results for $L\leq 30$ and zero temperature quantum Monte Carlo for $32\leq L \leq 
80$.  The agreement is remarkably good considering the severe difficulties of the 
extrapolation due to the logarithmic nature of the corrections.  In the remainder of this 
report we consider the nature of the corrections to this formula, for the Heisenberg 
antiferromagnet.

Let us begin with $ G_s(r)$ for an infinite system. The integral in the exponent in Eq. 
(\ref{corr}) can be rewritten as:
\begin{eqnarray} \int_{g_0}^{g(r)}[\gamma (g)/\beta (g)]dg&=&(1/2)\ln (r/r_0)+
\int_{g_0}^{g(r)}\left[{1\over 4g}+\sum_{n=0}^\infty a_ng^n\right] \nonumber \\
&=&(1/2)\ln (r/r_0)+(1/4)\ln [g(r)/g_0]+\sum_{n=0}^\infty {a_n\over n+1}
[g(r)^{n+1}-g_0^{n+1}].\end{eqnarray}
Here the $a_n$ terms arise from the higher order terms in the perturbative expansions 
of $\beta (g)$ and $\gamma (g)$.  Noting that  all terms involving $g_0$ are just 
constants, and also Taylor expanding the function $F[g(r)]$ in Eq. (\ref{corr}), we may 
finally write:
\begin{equation}  G_s(r)\to {1 \over r\sqrt{g(r)}}\sum_{n=0}^\infty 
b_ng^n(r),\end{equation}
in terms of some combined coefficients, $b_n$.  Including the cubic term in the 
$\beta$-function for the isotropic case\cite{Solyom}:
\begin{equation} 
dg/d\ln r = -(4\pi /\sqrt{3})g^2-(1/2)(4\pi /\sqrt{3})^2g^3.\end{equation}
Integrating gives:
\begin{equation} {1\over g(r)}-{1\over g_0}=(4\pi /\sqrt{3})\{\ln (r/r_0)+(1/2 )\ln [\ln 
(r/r_0)]\}+ O(1).\label{g(r)}
\end{equation}
Thus, we may write:
\begin{equation}
 G_s(r)={1 \over (2\pi )^{3/2}}{\{\ln (r/r_0)+(1/2)\ln [\ln (r/r_0)]\}^{1/2}\over r}[1+ 
O(1/\ln r)].\end{equation}
We may absorb the leading correction into a constant term inside the square root:
\begin{equation}
 G_s(r)={ 1\over (2\pi )^{3/2}}{\{\ln (Cr/r_0)+(1/2)\ln [\ln (r/r_0)]\}^{1/2}\over r}\{1+ 
O[1/(\ln r )^2]\}.\label{lnln}\end{equation}
From Eq. (\ref{g(r)}), $C$ has the form:
\begin{equation} C=e^{\sqrt{3}/4\pi g_0+ O(1)}.\label{C}\end{equation}
The O(1) term in the exponent in Eq. (\ref{C}) could be computed.  It requires 
calculation of the anomalous dimension $\gamma (g)$ to $O(g^2)$ and of the function 
$F$ to O(g).  This term was ignored in \cite{Hallberg} leading to an inaccurate 
determination of $g_0$.  

Let us now consider the Green's function on a ring of length L, $ G_s(r,r/L,g)$.  The 
RG equation, Eq. (\ref{RG}), is still obeyed.  The derivative in this equation may be 
taken either with respect to r or L with the ration r/L held fixed.  This follows because a 
rescaling of {\it both} length scales is equivalent to a coupling constant redefinition.
Using an L-derivative, the solution  is now:
\begin{equation}  G_s(r,L, g^0)=\exp \{-2\int_{r_0}^L d\ln r' \gamma [g( r')]\}F[g(L), 
r/L],\label{corrL}\end{equation}
The exponential factor is independent of $r$.  The function $F[ g(L), r/L]$ may be 
expanded perturbatively in $g(L)$ for large $L$:
\begin{equation}
F[ g(L), r/L]=\sum_{n=0}^\infty g(L)^nF_n(r/L).\label{sum}\end{equation}
The various functions $F_n(r/L)$ can be calculated by doing perturbation theory in the 
system with finite length.  They should all obey the periodicity requirement:
\begin{equation}F_n[r/L]=F_n[(L-r)/L].\end{equation}
If we take the asymptotic limit $r/L\to 0$, we should recover the infinite L result of Eq. 
(\ref{lnln}).  The zeroth order term, $F_0(r/L)$ is obtained by ignoring the marginal 
interaction altogether and simply calculating:
\begin{equation} <\hbox{tr}(\vec \sigma g)(r)\cdot \hbox{tr}(\vec \sigma 
g)(0)>_L\end{equation}
in the conformally invariant WZW model, on a circle of length L.  The correlation 
function on the circle (i.e. the cylinder in the space-time picture) is simply obtained by 
a conformal transformation and is given by:
\begin{equation} 
<\hbox{tr}(\vec \sigma g)(r)\cdot \hbox{tr}(\vec \sigma g)(0)>_L
\propto {1\over L\sin (\pi r/L)}.\end{equation}
Thus we may write:
\begin{equation}
 G_s(r,L)\to {1 \over (2\pi )^{3/2}}{\{\ln (L/r_0)+(1/2)\ln [\ln (L/r_0)]\}^{1/2}\over 
(L/\pi )\sin (\pi r/L)}\left[1+ {1\over \ln (L/r_0)}\tilde F_1(r/L)+ \ldots 
\right],\label{FSS}\end{equation}
for some other scaling function, $\tilde F_1$.  Alternatively, solving the RG equation 
with an r-derivative, we obtaine this result with L replaced by r inside all logarithms 
and a different scaling function $F_1'(r/L)$. [Note that, taking $r>>r_0$ with $r/L$ 
held fixed, the difference between $\sqrt{\ln (r/r_0)}$ and $\sqrt{\ln (L/r_0)}$ is 
suppressed by  a factor of $1/\ln (r/r_0)$.]

For the general xxz model the leading order finite-size scaling result is again obtained 
by the simple replacement:
\begin{equation} r\to (L/\pi )\sin (\pi r/L).\end{equation}
In particular, for $ G_s^x$ in the xx model ($\gamma =0$) we obtain:
\begin{equation}  G_s^x\propto [\sin (\pi r/L)]^{-1/2}.\end{equation}
The corrections are  down by powers of 1/r rather than only logarithms.

\begin{figure}
\epsfxsize=10 cm
\epsfbox[20 120 500 450]{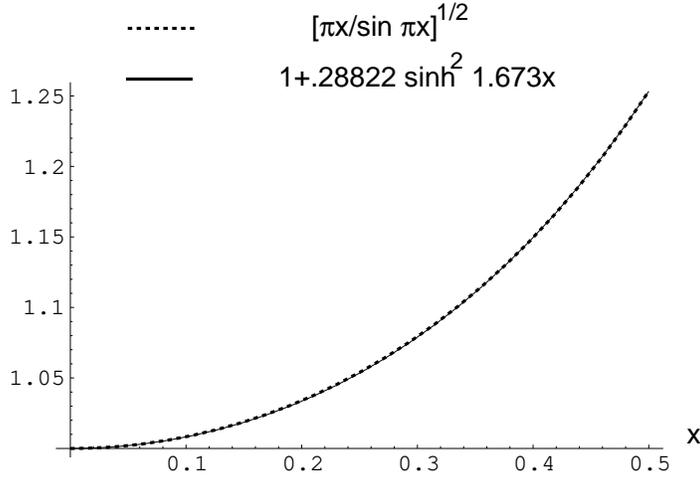}
\caption{Comparison of the 2 different scaling functions for the xx model.}
\label{fig:xx}
\end{figure}

\begin{figure}
\epsfxsize10 cm
\epsfbox[20 120 500 500]{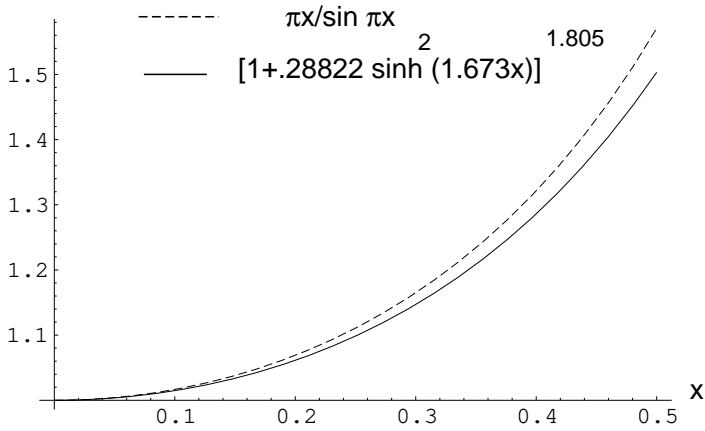}
\caption{Comparison of the 2 different scaling functions
for the xxx model.}
\label{fig:xxx}
\end{figure}

The efforts to fit numerical results on correlation functions in S=1/2 antiferromagets to 
a finite-size scaling form have a rather curious history. 
The case of $ G_s^x$ for the xx model was considered in \cite{Kubo}.  Rather than 
using the result predicted by conformal invariance the authors adopted a 
phenomenological expression, with free parameters adjusted to obtain good data 
collapse, corresponding to the replacement:
\begin{equation} \left( {\pi x\over \sin \pi x}\right)^{1/2}\to  
1+.28822 \sinh^2(1.673x),\end{equation} for $x\equiv r/L$.
This leads to a correlation function not obeying the periodicity condition:
\begin{equation}  G(r,L)= G(L-r,L).\end{equation}
Thus, the data fitting was only done for $0<x<1/2$.  Over this range, these two 
functions actually agree to within about .05\% as indicated in Figure \ref{fig:xx}.  This 
indicates that the conformal field theory (CFT) prediction is extremely accurate for the 
xx model.  It was proposed in \cite{Kubo} that, in the general xxz model, one should 
use the form:
\begin{equation}
[1+.28822 \sinh^2(1.673x)]^{2\eta},\label{wrong}\end{equation}
for $ G_s^x$.  This is essentially the correct CFT prediction, due to the numerical 
agreement noted above.  However, in \cite{Hallberg} the exponent in Eq. (\ref{wrong}) 
was taken to be a free parameter.  For the Heisenberg model a best fit was obtained 
with the exponent $1.805$ rather than the correct value of 2.  Thus the scaling form 
used differed slightly from the one predicted by CFT as shown in Fig. \ref{fig:xxx}.  
The maximum disagreement, at $x=.5$, is about 4\%.  

Koma and Mizukoshi used the scaling function
\begin{equation}
 G_s(r,L)\to {A\{\ln [(L/\pi r_0)\sin (\pi r/L)]\}^{1/2}\over (L/\pi )\sin (\pi 
r/L)},\label{KM}\end{equation}
obtaining a best fit for $A\approx .065$ (close to $(2\pi )^{-3/2}\approx .0635$). The 
agreement between this formula and their numerical data is better than 1.26\% for 
$1\leq r\leq L/2$ and $4\leq L \leq 80$.   Taylor expanding in $1/\ln (L/r_0)$, we see 
that this expression is consistent with Eq. (\ref{FSS}) for a particular choice of the 
function $\tilde F_1$, up to the small discrepancy in the amplitude.  Eq. (\ref{KM}) has 
the great advantage of simultaneously having the correct periodicity property and the 
correct behaviour in the limit $L\to \infty$.  However, such an expression can only arise 
from Eq. (\ref{corrL}) by summing an infinite number of terms in Eq. (\ref{sum}) [and 
ignoring the $\ln [\ln (L/r_0)]$ terms in $g(L)$].

 We expect that the somewhat larger discrepancy with CFT for the Heisenberg model 
than for the xx model can be accounted for by the log corrections.  The range of r used 
in the numerical work of Hallberg et al. \cite{Hallberg} for which fairly good data 
collapse was obtained was only $10<r<30$.  In this range we might expect the factor 
$1/\ln (r/r_0)$ in Eq. (\ref{FSS}) (written with r replaced by L inside the logarithms) to 
be fairly constant.  Thus the $F_1'$ term acts essentially as a small correction to the 
scaling function, $\pi x/\sin (\pi x)$.  (A related observation was made in \cite{Koma}.)  
It is feasible to push this renormalization group improved perturbation theory to one 
higher order and calculate $\tilde F_1(r/L)$ in Eq. (\ref{FSS}).  This involves using the 
known result for the $\beta$-function to $O(g^3)$, calculating the anomalous 
dimension to $O(g^2)$ and calculating the Green's function on a finite strip to $O(g)$.  
We expect that this could give better agreement with the numerical results and could, in 
particular,  reduce the small discrepancy between the exact amplitude and the results of 
\cite {Hallberg} and \cite{Koma}.  
I would like to thank M. Oshikawa for very helpful discussions and S. 
Lukyanov for informing me of his work.  After this paper was completed I encountered 
the  recent preprint \cite{Lukyanov2} which has some overlap with the results derived 
here.
This research  was supported by NSERC of Canada.  

\end{document}